# Continuously-tunable light–matter coupling in optical microcavities with 2D semiconductors

Franziska Wall, Oliver Mey, Lorenz Maximilian Schneider and Arash Rahimi-Iman*

*Faculty of Physics and Materials Sciences Center, Philipps-Universität Marburg, D-35032 Marburg, Germany*
*email: a.r-i@physik.uni-marburg.de*

**Abstract:**

A theoretical variation between the two distinct light–matter coupling regimes, namely weak and strong coupling, becomes uniquely feasible in open optical Fabry-Pérot microcavities with low mode volume, as discussed here. In combination with monolayers of transition-metal dichalcogenides (TMDCs) such as $WS_2$, which exhibits a large exciton oscillator strength and binding energy, the room-temperature observation of hybrid bosonic quasiparticles, referred to as exciton–polaritons and characterized by a Rabi splitting, comes into reach. In this context, our simulations using the transfer-matrix method show how to tailor and alter the coupling strength actively by varying the relative field strength at the excitons' position – exploiting a tunable cavity length, a transparent PMMA spacer layer and angle-dependencies of optical resonances. Continuously tunable coupling for future experiments is hereby proposed, capable of real-time adjustable Rabi splitting as well as switching between the two coupling regimes. Being nearly independent of the chosen material, the suggested structure could also be used in the context of light–matter-coupling experiments with quantum dots, molecules or quantum wells. While the adjustable polariton energy levels could be utilized for polariton-chemistry or optical sensing, cavities that allow working at the exceptional point promise the exploration of topological properties of that point.

## Introduction: Tunable light–matter interaction

Light–matter-interaction in optical microcavities has been the subject of nearly three decades of research, involving all kinds of quantum emitters and cavity types [1-3]. The coupling of excited matter states and photon modes can be characterized by two regimes. The weak-coupling regime, leading to altered radiative excitation lifetimes and enhanced or suppressed spontaneous emission due to the Purcell effect [4], and the strong-coupling regime. This regime is manifested by a normal mode splitting — known as (vacuum) Rabi splitting — and coherent reversible energy exchange between light and matter states, ultimately leading to the formation of hybrid bosonic quasiparticles [5, 6]. A requirement is that the coupling strength is high enough and decay rates sufficiently low. Besides quantum information and light-source developments on the application side [3] it is the field of fundamental cavity quantum electrodynamics studies [7] which demands optical high-quality microcavities. The topic highly benefits from research on various confinement concepts and has strongly driven the overall developments [8-10].

Since their discovery by Weisbuch in 1992, exciton-polaritons in optical microcavities, also known as cavity-polaritons, have attracted considerable attention by the research community, particularly for the first experimental realization of a Bose-Einstein condensate (BEC) [11, 12] and later for the first achievement of novel practical devices [13, 14], for room temperature condensate [15, 16] as well as polariton-lasing [17, 18] demonstrations, owing to the light effective mass of the hybrid bosonic quasiparticles. In this context, two-dimensional semiconductors such as $WS_2$ and $WSe_2$ from the family of transition-metal dichalcogenides are currently strongly under consideration for the next material system in optical cavities [19-24] with which practical room-temperature BEC experiments could be carried out.

For Fabry-Pérot microcavities, several parameters such as the mirror distance in combination with the angle of incidence for the incoming light or the emitter position inside of the cavity provide the possibility of continuous tuning of the coupling situation in light–matter interactions. In this work, we propose two ideas how to tailor the coupling strength by the relative field strength at the exciton position. Firstly, a more intuitive approach is considered, i.e. displacing a 2D material with respect to the resonator field inside an open planar microcavity with fixed mirror separation. Secondly, based on this, an open cavity with extra spacing layer between 2D material and one mirror is introduced. This alternative setup provides the advantage that it consists only of two independent movable parts and is therefore presumably easier to realize in an experimental setup. Furthermore it is independent from the choice of the active material. The cavity length is changed to set the excitons' position within the standing-wave field pattern. At the same time, spectral overlap between cavity and exciton modes occurs at different angles of incidence for the incoming cavity light. We have analyzed this structure configuration by calculating the angle-resolved reflectivity spectra as well as the field distribution for a monolayer WS$_2$ using the transfer-matrix method. Thereby, we show that a transition between the weak and strong coupling regime can be observed. The coupling strength is completely tunable in our proposed setup by the choice of the spacer layer's thickness. These findings could enable the exploration of the transition between weak and strong coupling – especially the exceptional point – by prospective actively tunable experiments. Furthermore, the material independent continuous adjustability could be utilized for polariton-based chemistry allowing a precise adjustment of the polariton energy levels [25]. This could provide an experimental stage for an efficient study of chemical reactions influenced by different light–matter coupling situations.

## Theoretical Background

The Rabi splitting is the figure of merit in strong-coupling experiments and is written for energetic resonance between cavity and exciton as $\hbar\Omega_{Rabi} = E_{UP} - E_{LP} = 2g_0$, whereas $E_{UP}$ and $E_{LP}$ are the resulting energy levels of the upper and lower exciton–polariton modes, respectively. For clarity, the momentum-space dependencies of the modes, which lead to angle-dependent emission properties, are blanked out for a moment in our discussion, while coupling does not violate energy and momentum conservation.

The maximal coupling strength $g_0$ is determined, besides the physical constants for the elementary charge $e$, the vacuum permittivity $\epsilon_0$ and the electron mass $m_e$, by the cavity's permittivity $\epsilon$, the exciton's oscillator strength $f$ as well as the cavity's mode volume $V_{eff}$,

$$g_0 = \sqrt{\frac{1}{4\pi\epsilon\epsilon_0}\frac{\pi e^2 f}{m_e V_{eff}}} \tag{1}$$

In general, the physical coupling strength $g$ is described by the maximal possible coupling strength $g_0$, the relative field strength reflecting the position inside the field pattern $\psi(z)$ and the angle between the electric field and the exciton's dipole moment $\xi$ [2]. This accounts for the anisotropy of WS$_2$. An explicit consideration of the out-of-plane dark exciton is not necessary, because of its weak oscillator strength [26, 27] as discussed in the supplement.

$$g = g_0 \psi(z) cos(\xi) \tag{2}$$

$$\psi(z) = \frac{E(z)}{|E_{max}|} \tag{3}$$

The light-matter-coupling is modelled by the following Hamiltonian.

$$\hat{H} = \hbar(\omega_{cav} - \frac{i\gamma_{cav}}{2})a^\dagger a + \hbar(\omega_{ex} - \frac{i\gamma_{ex}}{2})\sigma^\dagger\sigma + \hbar g(\sigma^\dagger a + a^\dagger\sigma) \tag{4}$$

with the creation and annihilation operators of both species a†/σ† and a/σ, respectively. The first term describes the cavity field with its resonance frequency $\omega_{cav}$, the second one the exciton with its resonance frequency $\omega_{ex}$ and the last one the coupling between both.

Based on this, the eigenvalues represent the resonance frequencies of the coupled system,

$$\omega_{\pm} = \frac{1}{2}\left(\omega_{ex} + \omega_{cav} + \frac{i(\gamma_{ex}+\gamma_{cav})}{2}\right) \pm \frac{1}{2}\sqrt{(2g)^2 + \left(\omega_{cav} - \omega_{ex} + i\frac{(\gamma_{cav}-\gamma_{ex})}{2}\right)^2} \tag{5}$$

The coupling regime is mainly given by the coupling strength as well as the cavity's linewidth $\gamma_{cav}$ and the exciton's linewidth $\gamma_{ex}$. In the literature, often two criteria appear: $4g^2 > |\gamma_{cav} - \gamma_{ex}|$ as condition for a splitting of the eigenvalues in Eq. (1) (e.g. in [28]) and $4g^2 > |\gamma_{cav} + \gamma_{ex}|$ as requirement for a resolvable splitting, for which the frequency splitting is larger than the polariton linewidth (c.f. [24, 29]). By tuning the coupling, it becomes possible to especially investigate the transition between weak and strong coupling. Between the two regimes lies an exceptional point (EP). At this position, the coupled system only features one complex solution [28, 30, 31]. This gives rise to a variety of physical phenomena like chiral behaviour [32] as well as an energy transfer between the two modes of the strongly coupled system [33] under encirclement of the EP. Due to the sharp transition between the coupling regimes, an EP is also suitable for the design of sensors [34].

In order to tune from one to another coupling regime, either the excitons' or the cavity's properties can be altered. Tunable coupling has been demonstrated with WS$_2$ by applying a current [35] or varying the temperature [36], causing a change of the exciton oscillator strength. Furthermore, chemical reactions of molecules can be used to switch between the coupling regimes [37]. Moreover, in monolithic cavities, the dependency of the coupling situation on the relative field strength has been demonstrated for molecules [38]. Recently, continuously tunable coupling was shown for carbon-nano tubes by utilizing the polarization of the light [39].

All mentioned methods above either depend on the choice of the material, need multiple growth processes or are not actively tunable over a wide range. Here, we use properties of an open cavity to continuously tune the coupling strength and thereby are giving the perspective of real-time switching between the strong and weak coupling regime without influencing the active region. Besides this, the presented method keeps the spectral position of the cavity mode resonant to the exciton. Our suggested structure is *a priori* not limited to a special material. For the active medium providing the excitons only a high enough oscillator strength to reach strong coupling as well as a sub-wavelength-thickness compared to the emission wavelength is required. With the aim in mind to control the coupling situation without changing the setup in between experiments, we chose a planar open microcavity. In combination with monolayers of transition-metal dichalcogenides such as WS$_2$, the excitons of which exhibit a large oscillator strength, the strong coupling regime can be reached with an open cavity already at room temperature [20].

## 2D–Microcavity Structure

All simulations are done for s-polarized light in order to exclude the influence of a misalignment between the dipole moment of excitons and the cavity mode's electric field and thereby to ensure that they are perpendicular to each other for every angle of incidence. The simulated Fabry-Pérot (FP) cavity consists of two dielectric mirrors (DBR), which comprise alternating layers of SiO$_2$ and Si$_3$N$_4$, with their wavelength dependent refractive indices extracted from [40] and [41], respectively. A similar configuration was already presented in the literature to show room-temperature polaritons for WS$_2$ [20]. The high-reflectivity mirrors end with the low refractive index material, namely SiO$_2$, to ensure the field maximum at the DBRs' surface for normal incidence. Moreover, directional leakage of cavity photons is enabled by one side featuring less mirror pairs, resulting in a lower reflectivity for the corresponding mirror. The two mirrors therefore consist of 11.5 and 12.5 $SiO_2/Si_3N_4$ layer pairs. Acting as an active medium, a monolayer of the transition-metal dichalcogenide $WS_2$ is – in the simulations – brought into coupling with the cavity mode of the open micro-resonator. Its optical behavior is modeled by an effective optical thickness of 6.18 Å [42]. The wavelength-dependent complex refractive index is derived from experimental data (from Ref. [42]) together with an approximation of that data which consists of a summation over Lorentzian peaks according to Ref. [43] (see Supporting Information Fig. SI. 1). A scheme of the simulated structure is shown in Fig. 1a.

For the spacer layer in our second tunable-coupling approach, we simulated PMMA on top of one mirror (Fig. 1b). This material is already used in waveguide applications [44] or as spacer layer in monolithic cavities [38] due to its transparency for visible light [45]. Further, it enables a wide range of possible thicknesses, from the nm-range up to several 100 nm. The tunable coupling is achieved by variation of the angle of incidence for the incoming light combined with an adjustable cavity length. Thus, we simulated angle-resolved reflectivity spectra to identify the coupling situation.

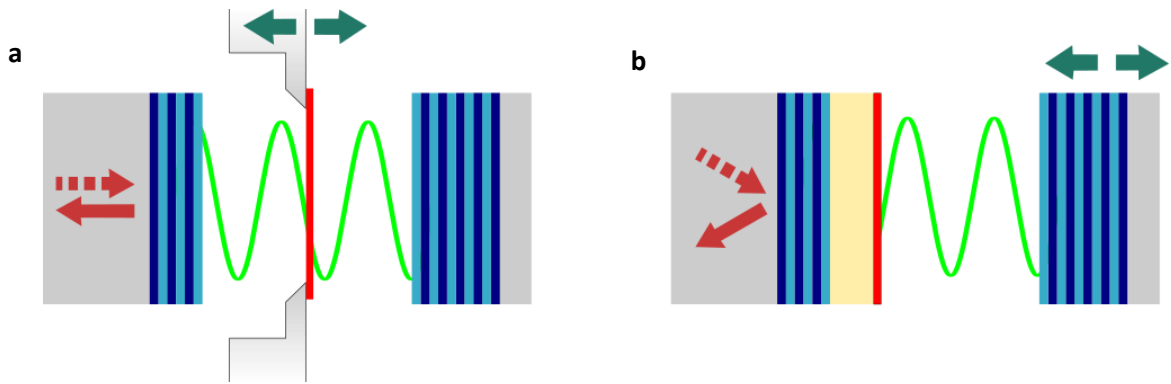

Fig. 1: Schematic drawing representing a tunable-cavity coupling approach for (a) a structure without spacer layer incorporating a free-standing monolayer and (b) a structure with intra-cavity spacer layer, on which a monolayer is placed.

## Simulations and Discussion

### *Variation of the 2D-material's position*

As an intuitive approach, the excitons' position with respect to the field pattern $z$ can be varied by shifting the material inside of an open microcavity with fixed cavity length. As a result of the adjusted position, the Rabi splitting $\hbar\Omega_{Rabi}$ can be varied, which can lead to a change in the coupling regime. Figure 2a displays a plot of the system's resonance frequencies based on the calculated reflectivity spectra of the complete microresonator structure. At positions in the cavity corresponding to a strong coupling regime, characterized by a mode splitting, an asymmetric energy splitting into a photonic lower polariton branch (blue squares) and an excitonic upper polariton branch (red triangles) around the exciton's energy (dashed horizontal line) is noticeable. This is a hint, that the system is not yet completely resonant. Here, the cavity length is determined in a way, that the resonance energy of the empty cavity without 2D-material overlaps with the bare exciton mode (dashed horizontal line), which is estimated at a wavelength of 617.2 nm (rounded) by the imaginary part of the refractive index representing the absorption. In Fig. 2a the solid black curve represents a cavity mode with $WS_2$ at different spatial positions for which the imaginary part of the 2D material's refractive index is neglected. Consequently, only the reflectivity of $WS_2$ is included for the solid black line but not its absorption. The simulated curve indicates that the 2D material's reflectivity already detunes the spectral position of the cavity mode without coupling considerations. This can result in a position dependent energetic offset between photon and exciton mode within the final setup. The cavity's resonance has no offset with respect to the exciton mode at the minima of the relative field strength ($z \approx 0.15\mu m$). At this position, the cavity is not coupled to the excitonic mode. It is worth to keep in mind, that the empty uncoupled-cavity mode has not the same spectral position as the calculated cavity for a system with active region. Beside this unreachable zero-detuning configuration for each monolayer position, this approach has from the experimental point of view also the drawback of requiring a complex, rigid and perforated holder to incorporate the free-standing material into the resonator. The calculated energy splitting normalized to the maximum (shown as dots in Fig. 2b) is nearly proportional to the theoretical curve representing the absolute electric field strength of this cavity configuration (black line). For clarity, also here the refractive index modulation of the structure is displayed in the background.

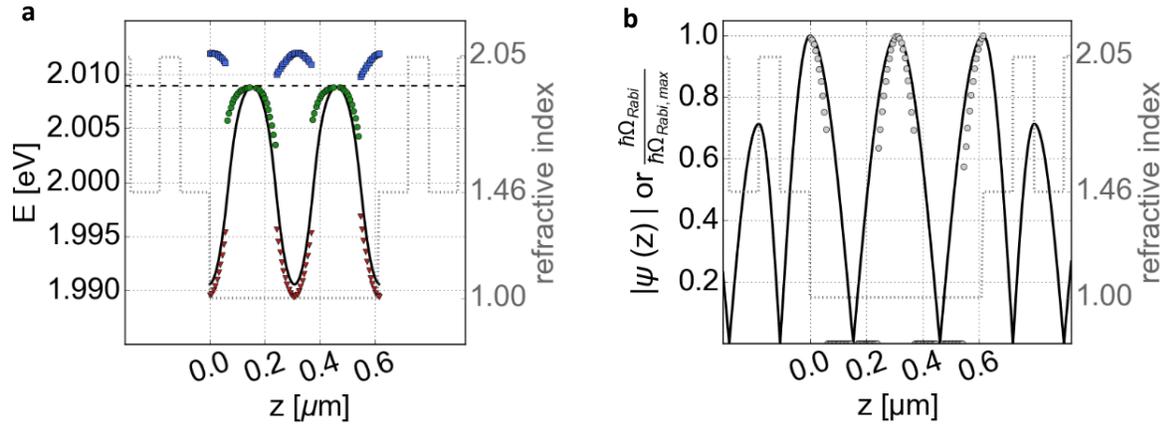

Fig. 2. (a) Eigen-energies of the upper and lower polariton states (blue squares and red triangles, respectively) as well as of the weakly coupled cavity system (green circles) based on calculated reflectivity spectra (corresponding to a setup like in Fig. 1a). For each simulation, the 2D material WS$_2$ is placed at the corresponding distance $z$ towards the left mirror. The black dashed line marks the exciton's energy (left scale). The solid black line represents the cavity's mode for which the imaginary part of the refractive index of WS$_2$ is neglected (corresponding to an uncoupled system, i.e. E$_{res}$ = ℏω$_{cav}$). (b) Calculated absolute relative electric field strength of the empty cavity (black line) with dots representing the calculated coupled-mode splitting normalized to its maximum (left axis). In all graphs, the gray dotted line in the background marks the refractive index modulation (right scale) of the empty cavity consisting of DBRs and an air gap.

### Concept making use of an additional spacing-layer

The optical phase shift $\varphi_{cav}$ between two plane parallel mirrors depends on the geometric path difference $\Delta s_{cav}$, given by the angle of incidence $\theta$ for the incoming light, the refractive index $n$ and the cavity length $L_{cav}$, as well as the wavelength $\lambda$.

$$\varphi_{cav} = \frac{2\pi \Delta s_{cav}}{\lambda} = 2\pi \frac{(n\cos(\vartheta)2L_{cav})}{\lambda} = 2\pi \frac{(n\cos(\vartheta)2L_{cav})\,E_{cav}}{h\,c} \ . \tag{6}$$

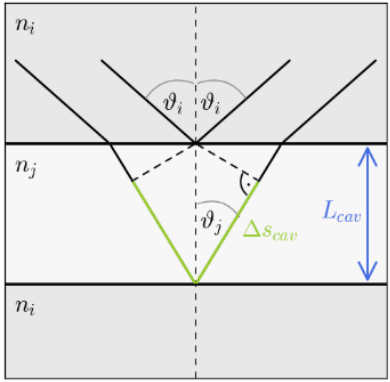

Fig. 3: Optical path between two plane parallel mirrors.

For the resonant FP cavity mode, $\varphi_{cav}$ is a multiple of $2\pi$. Waves with another wavelength $\lambda$ for which $\varphi_{cav}$ is not a multiple of $2\pi$ interfere completely or partly destructively. If the angle of incidence $\vartheta$ is increased, the path difference and thereby the phase shift decrease. Consequently, the wavelength of the cavity mode is reduced and the energy of the (uncoupled) cavity mode $E_{cav}$ is increased (see Eq. (6)). For a longer cavity the path difference is increased leading to a reduction of the cavity mode's energy. Hence, the influence of $\theta$ on $E_{cav}$ could be compensated by changing $L_{cav}$ (see Fig. 3). By adding an additional spacer layer, the optical path difference between the two mirror surfaces gets the sum of the path difference inside of the PMMA-layer $\varphi_{PMMA}$ and inside of the air gap $\varphi_{Air}$,

$$\varphi_{cav} = \varphi_{PMMA} + \varphi_{Air} = \frac{2\pi}{\lambda} 2(n_{PMMA}\cos(\vartheta)\mathrm{d}_{PMMA} + n_{Air}\cos(\vartheta)\mathrm{d}_{Air}) \tag{7}$$

For simplicity, the influence of the mirrors reflectivity on the path difference is neglected here. Increasing $\vartheta$ shortens the geometric path within the air gap $\Delta s_{Air}$ as well as the PMMA layer $\Delta s_{PMMA}$. Thus, the corresponding phase shifts and thereby $\varphi_{cav}$ are decreased. Consequently, $E_{cav}$ is increased. Again, this could be in principle compensated by changing $L_{cav}$. The cavity length is given by the layer thicknesses, $L_{cav} = \mathrm{d}_{PMMA} + d_{Air}$. Here, only the gap size $d_{Air}$ is adjustable, while $\mathrm{d}_{PMMA}$ stays constant. Increasing the mirror separation leads to an increased gap size $d_{Air}$. Accordingly, $\varphi_{cav}$ is increased and $E_{cav}$ is decreased. Hence, also an increased gap size could be used to compensate the phase shift induced by an increased $\vartheta$. In summary, increasing the cavity length leads to a spectral overlap between exciton resonance and cavity mode at higher angles of incidence. Here, this angle is denoted as resonant angle $\vartheta_{res}$. In the following, we consider only the case of spectral resonance (i.e. zero-detuning: $E_{cav}(\vartheta_{res})$ - $E_{ex}$= $0$) and thereby each change in the cavity length corresponds to a change in $\vartheta_{res}$. The cavity mode's spectral position is determined by including the monolayer's reflectivity while neglecting its absorption to exclude the detuning observed for the structure in Fig 1a (see Supplement Fig.SI. 6). Again, the exciton wavelength in our simulation is estimated as rounded to 617.2 nm by the peak position within the imaginary part of the dielectric function.

However, $\varphi_{PMMA}$ and $\varphi_{Air}$ are now changed in a different manner, although the overall phase shift $\varphi_{cav}$ stays the same. Increasing $\vartheta$ causes a decreased $\varphi_{PMMA}$ as well as $\varphi_{Air}$. At the same time, the additionally increased $L_{cav}$ leads to an increased phase shift only within the air layer. Due to the required spectral resonance between exciton and cavity, the total phase shift $\varphi_{cav}$ of the structure is kept constant. In sum, $\varphi_{PMMA}$ is reduced whereas $\varphi_{Air}$ is increased. Together with this mismatch between the change of phase shift inside PMMA and air, it gives rise to a shift of the field pattern relative to the PMMA–air interface. The phase shift $\varphi_{PMMA}$ determines the interface's position inside of the standing wave pattern and thereby the corresponding field amplitude. By placing the 2D material at this interface, these conditions can be utilized to adjust the coupling strength by tuning the relative field strength. This effect is illustrated within an additional video (see Supplementary Information), showing the field pattern as well as the angle-resolved reflectivity spectra for the suggested structure. For the field calculation the reflectivity of WS$_2$ is again considered while its absorption is neglected. Figure 4 shows three representative settings for such a structure (see Fig. 1b). The field pattern is moreover simulated for $\vartheta_{res}$. Different energy splittings can be obtained in the angle-resolved reflectivity spectra. Furthermore, the angle of incidence $\vartheta$ thereby not only alters the cavity mode's spectral position as well as the relative field strength but in general also the cavity's linewidth $\gamma_{cav}$, because modifying $\vartheta$ influences the reflectivity of the DBR mirrors. For this setup the change is less than 0.1% within the region of experimentally interesting angles (< 40°) and is thus not further discussed as impact on the coupling regime. Nevertheless, the mirror contributes an additional phase shift for higher angle of incidences (See Supplement Fig. SI 7). Thus, spectral resonance is determined by iteratively simulating the reflectivity spectra for varied mirror separations. However, the reflection dip in the lower row within the last image is slightly shifted with respect to the exciton energy indicating an influence of the monolayer's absorption on the cavity's spectral position.

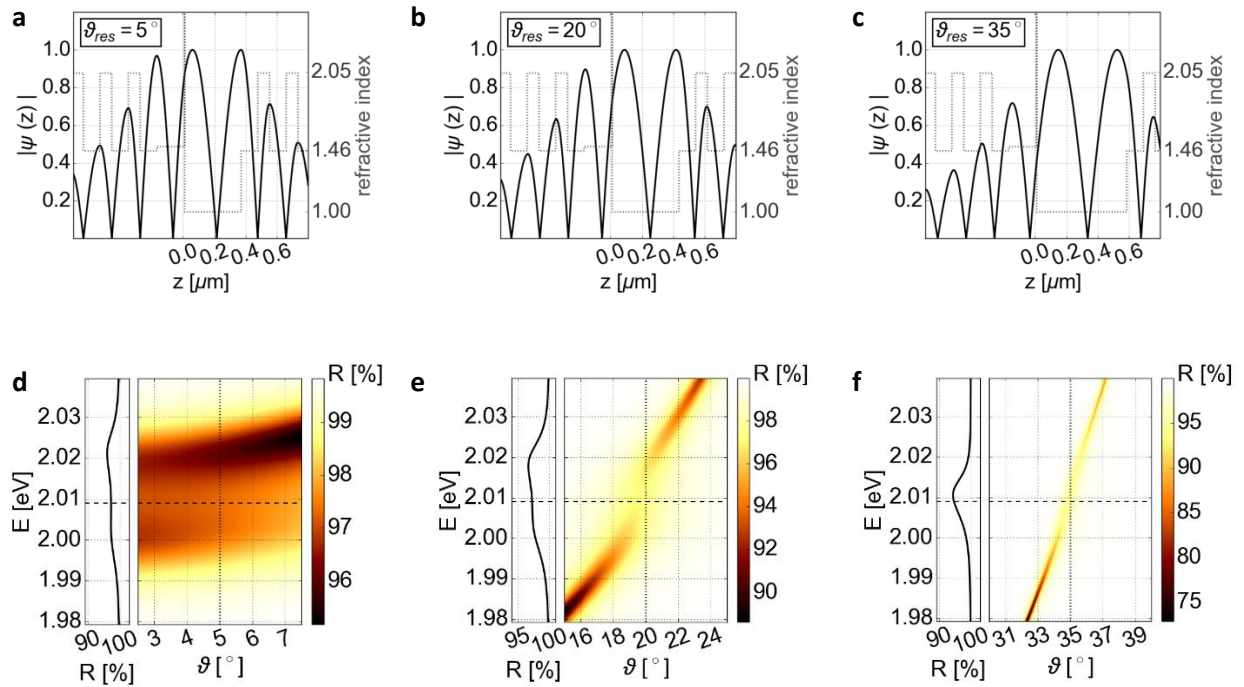

Fig. 4: Impact of a changed cavity length on an open planar microcavity with a 178.2 nm thick PMMA spacer layer on one mirror (structure depicted in Fig.1b). The second cavity mode corresponds from left to right to an air gap of 360.8 nm, 425.6 nm and 575.7 nm, respectively. The upper row shows the simulated field distribution of an cavity at resonance ($E_{cav}$ ($\vartheta_{res}$) - $E_{ex}$= 0) , which is marked in the angle-resolved reflectivity spectra below by a dashed vertical line, with neglected absorption of WS$_2$ while the horizontal dashed line indicates the exciton's emission energy. The exciton's spectral position is determined by the peak position within the imaginary refractive index. The influence of the first three cavity modes could be found in the Supporting Information (Fig. SI. 4 and 5). Further, the corresponding transmission and absorption spectra as well as the spectra for p-polarized light are displayed within the Supplement (Fig. SI. 8 and Fig. SI. 9).

***Additionally tailoring the coupling situation***

The spacer layer's thickness $d_{PMMA}$ defines the starting position in the standing wave pattern at normal incidence by the corresponding phase shift $\varphi_{PMMA}$. At normal incidence, different air gaps are necessary for different PMMA thicknesses to reach spectral resonance. Therefore, the change in cavity length $\Delta z$ with respect to $L_{cav}(\vartheta_{res} = 0)$ is used in Fig. 5 as a common axis ($\Delta z = L_{cav}(\vartheta_{res}) - L_{cav}(\vartheta_{res} = 0)$). For decreasing cavity length compared to $L_{cav}(\vartheta_{res} = 0)$, the energetic resonance condition can no longer be reached, because a further increase of $\cos(\vartheta)$ in Eq. (7) in order to compensate the phase shift compared to $\vartheta = 0$ is not possible. The simulated relationship between $\Delta z$ and $\vartheta_{res}$ is displayed within Fig. 5a. On the other hand, increasing the cavity length shifts the exciton's location relative to the field pattern towards the left side of the field maximum (left direction in Fig. 4 a-c) due to the decreased $\varphi_{PMMA}$. By the choice of $d_{PMMA}$, it is possible to either reduce or increase the absolute field strength at the monolayer position for increasing $\vartheta$ as well as $\Delta z$ (see Fig. 5b). Thus, depending on the layer's thickness, different coupling scenarios can be reached.

In Fig. 5c, the Rabi splitting for different coupling situations based on various thicknesses of the spacer layer is shown. For example, if the PMMA–air-interface lies near the field maximum at normal incidence, a length tuning leads to a decreasing relative field strength and thereby coupling strength (e.g. blue and green curve in Fig. 5b for a 178.2 and 190.6nm PMMA layer). Similarly, increasing coupling under cavity length tuning can be reached, for instance, by a thickness $d_{PMMA}$ larger than one half of the wavelength in PMMA (e.g. 227.9 nm or 248.6 nm). A careful choice also enables the transition between the coupling regimes. Here, the exact position of the transition between the coupling regimes is not directly detectable. No splitting in this context means, that it is either not resolvable or the system is weakly coupled and thus indeed does not feature any resonance splitting. Nevertheless, the presented spacer-layer method allows a continuous tuning of the coupling strength (see supplementary video). Between no coupling at the field minimum and the maximal strong coupling at the field maximum should be a coupling strength referring to weak coupling prior to reaching the EP. Thus, weak coupling could be realized by placing the monolayer near the field minimum combined with the given tuning mechanism.

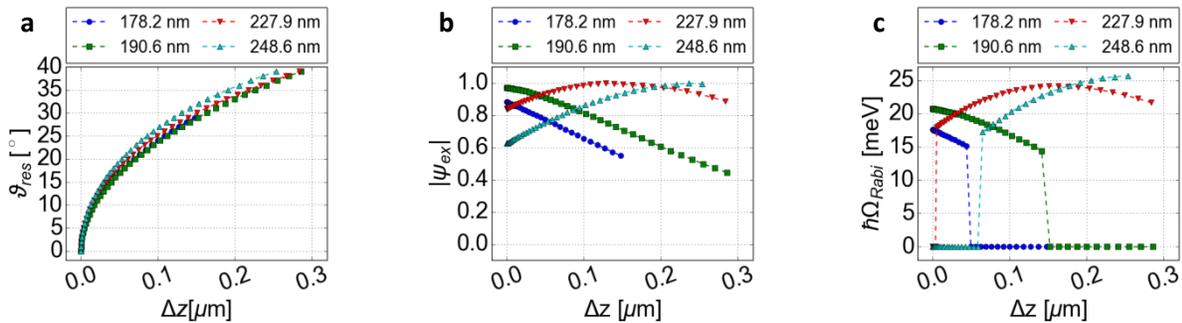

Fig. 5: (a) Relationship between the change in the cavity length $\Delta z$ and the resonant angle $\vartheta_{res}$ for different spacer layer's thicknesses $d_{PMMA}$ for the structure in Fig.1b. (b) Corresponding influence of $\Delta z$ on the relative field strength at the excitons' position $\psi_{ex}$ for $\vartheta_{res}$. The reflectivity of WS$_2$ is in those two calculations (a,b) considered, while its absorption is neglected. (c) Resulting resolvable splitting $\hbar\Omega_{Rabi}$ at $\vartheta_{res}$ obtained by a varied $\Delta z$ combined with a resonance adjustment by the angle of incidence $\vartheta$. The splitting is determined by the difference between the maxima of two Lorentzian peaks approximating the coupled modes fitted to the calculated reflectivity spectra at $E_{cav}(\vartheta_{res})$ - $E_{ex}$= 0.

# Conclusion

In summary, we have presented two designs for an open planar FP microcavity system, which allow nearly material independent, active and continuous modification of the light-matter interaction in the structure. Hence, such setups can be especially utilized as a testbed for real-time investigations on the transition between the strong and weak coupling regime. Beyond the employment of 2D-cavity systems for optoelectronic coupling, the alternative free-standing-sheet configuration could even enable sophisticated optomechanical experiments. The more practical variant utilizing a spacer layer on one of the DBRs of the FP cavity, as an alternative concept to a displaceable free-standing sheet medium, is characterized theoretically with a transfer-matrix model for a tuneable variation of the coupling strength. The choice of the spacer layer's thickness allows a precise control of the coupling situation. Thereby, an increase and decrease in the obtainable Rabi splitting as well as a modification of the coupling regime become possible scenarios while spectral resonance between cavity and exciton is realized. A setup of this kind is suitable for material systems with a sub-wavelength-thickness with respect to their exciton emission wavelength, provided that a large enough oscillator strength for strong coupling is exhibited. In the future, this method could be used to investigate light-matter coupling not only for 2D semiconductors, but also e.g. quantum dots, perovskite thin films, nanoplatelets or molecular films.

**Acknowledgement**

The authors acknowledge financial support by the German Research Foundation (DFG: SFB1083, RA2841/5-1), by the Philipps-Universität Marburg, by the Federal Ministry of Education and Research (BMBF) in the frame of the German Academic Exchange Service's (DAAD) program Strategic Partnerships and Thematic Networks. The authors thank Fang Wei from the Zhejiang University, China, for useful discussions.

**Authors' contributions**

A.R.-I. initiated the light–matter coupling considerations with tunable open microcavities and guided the work. F.W. contributed the spacing-layer idea, the implementation of the TMM software and performed the simulations. O.M. was involved in the development of the spacing-layer idea, the TMM calculation tool as well as the visualization. L.M.S. supported the overall research. The manuscript was written by all authors.

**Corresponding author**

Arash Rahimi-Iman: <u>a.r-i@physik.uni-marburg.de</u>

**Authors' statement/Competing interests**

The authors declare no conflict of interest

**Additional information**

Supplementary Information accompanies this paper

# Supporting Information (SI)

# Continuously-tunable light–matter coupling in optical microcavities with 2D semiconductors

Franziska Wall, Oliver Mey, Lorenz Maximilian Schneider and Arash Rahimi-Iman*

*Faculty of Physics and Materials Sciences Center, Philipps-Universität Marburg, D-35032 Marburg, Germany*
*email: a.r-i@physik.uni-marburg.de*

## Complex refractive index

In order to obtain the coupled-system's reflectivity spectra, the role of the transition-metal dichalcogenide monolayer inside the cavity is represented in our simulations by the complex refractive index extracted from the literature. Therefore, experimental data from Ref [42] is modelled in a reasonable approximation by a summation over multiple Lorentzian peaks according to Ref. [43].

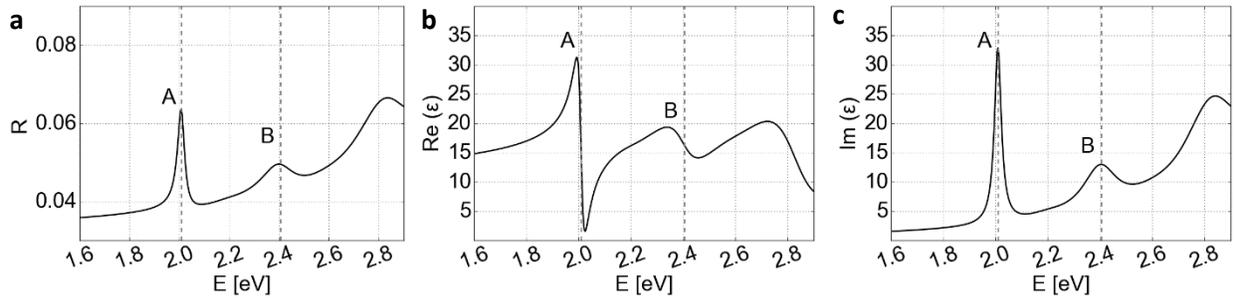

**Fig. SI. 1**: Simulated reflectivity of a WS₂ monolayer on top of an SiO₂ layer obtained by an approximation of experimental data from Ref. [40] with a summation over Lorentzian peaks according to Ref. [41] (a). Corresponding real and imaginary parts of the dielectric function, which are used here within the transfer-matrix method, are shown in (b) and (c), respectively. The vertical dashed lines mark the A and B exciton position.

Figure SI.1 shows the corresponding curve of the reflectivity (a) of a monolayer on top of silica with an optical thickness of 6.18 Å as well as the real (b) and imaginary part (c) of its dielectric function based on the data from [42] and [43]. Here, the exciton wavelength is determined by the peak position in the imaginary part of the dielectric function. The dependency of the dielectric function on the angle of incidence as well as on the light's polarisation is neglected in our simulation. For further investigations a detailed measurement of the angle-dependent reflectivity of a monolayer would be desired to get access to this information complemented by an angle-resolving theoretical modelling of the complex dielectric function. The in-plane permittivity $\epsilon_{||}$ of a WS₂ monolayer is larger than its out-of-plane $\epsilon_{\perp}$ [S1]. Thus, the resonant cavity length in out-of-plane direction is for $\vartheta = 90°$ roughly 1.44 times larger compared to the value estimated by the in-plane permittivity.

$$2\,\pi = \Delta s_{cav} = \sqrt{\epsilon}\,L_{cav}\cos(\vartheta), \tag{SI 1}$$

$$L_{cav,\perp} = \sqrt{\frac{\epsilon_{||}}{\epsilon_{\perp}}}L_{cav,||} \approx 1.44\,L_{cav,||}\,. \tag{SI 2}$$

Further, for s-polarized light the out-of-plane exciton cannot couple with the light field, because its dipole moment is perpendicular to the electric field. Moreover, the out-of-plane exciton has an about $10^3$ smaller oscillator strength compared to the in-plane exciton [S2]. Consequently, the coupling for out-of-plane components with p-polarized light is reduced by about 1.5 orders of magnitude compared to in-plane components. Thus, the effect of anisotropy is neglected in our considerations.

## Transfer-matrix method

Optical properties of thin-film stacks can be modelled one-dimensionally by the transfer-matrix method. Here, we give a short overview of the applied formulas in our simulation code. A detailed description could be found e.g. in [S3-S5]. Plane waves travelling in the $z$ direction perpendicular to the layer interfaces are described as superposition of one right propagating $E_n^- = 0$ and one left propagating $E_i^-$ wave in layer $i$, whereas each layer is assumed to be homogeneous.

$$E(z) = E_i^+ \exp(ik_{i,z}z) + E_i^- \exp(-ik_{i,z}z). \tag{SI 3}$$

Here, we chose the minus-sign inside of the exponential function for waves propagating to the structure's left side.

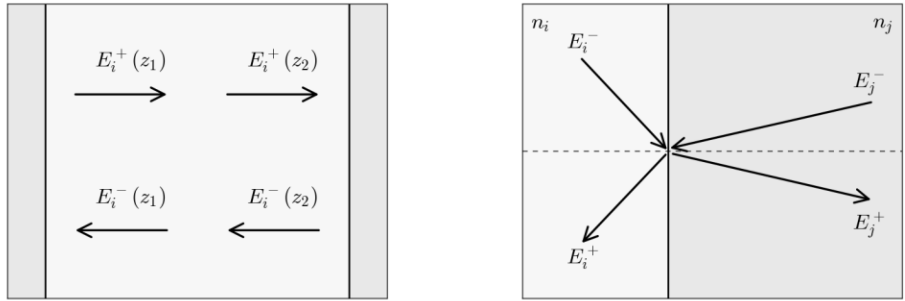

**Fig.SI. 2**: Schematic representation of the waves' propagation as well as behaviour at a layer interface which is used in the transfer matrix method.

The wave propagation inside of a layer $i$ is described by a phase shift $\varphi_i = 2\pi d_i \cos(\vartheta_i)/\lambda$, which is determined by the travelled distance $d_i$ and the wave vector $k_{i,z}$ in $z$ direction. The later one is given by the wavelength $\lambda$, the angle of incidence $\vartheta_i$ and the material's complex refractive index $n_i$.

$$E_i^+(z = d_i) = exp(i\varphi_i)E_i^+(z = 0), \tag{SI 4}$$

$$E_i^-(z = 0) = exp(i\varphi_i)E_i^-(z = d_i). \tag{SI 5}$$

Using a matrix formalism, the phase matrix $P_i$ is defined as following:

$$\begin{pmatrix} E_i^+(z = 0) \\ E_i^-(z = 0) \end{pmatrix} = \begin{pmatrix} exp(-i\varphi_i) & 0 \\ 0 & exp(i\varphi_i) \end{pmatrix} \begin{pmatrix} E_i^+(z = d_i) \\ E_i^-(z = d_i) \end{pmatrix} = P_i \begin{pmatrix} E_i^+(z = d_i) \\ E_i^-(z = d_i) \end{pmatrix}. \tag{SI 6}$$

A wave transmitted through the interface from layer $i$ to layer j can be modelled by a reflection coefficient $r_{ij}$ and transmission coefficient $t_{ij}$. The index' order represents thereby from which layer to which the wave is propagating.

$$E_j^+(z) = t_{ij}E_i^+(z) + r_{ji}E_j^-(z), \tag{SI 7}$$

$$E_j^-(z) = r_{ij}E_i^+(z) + t_{ji}E_j^-(z). \tag{SI 8}$$

The reflection and transmission coefficient are calculated by the Fresnel equations. In our case, the equation for s-polarized light is applied.

$$r_{ij,s} = \frac{E_i^-}{E_i^+} = \frac{n_i cos(\vartheta_i) - n_j cos(\vartheta_j)}{n_i cos(\vartheta_i) + n_j cos(\vartheta_j)} \ , \tag{SI 9}$$

$$t_{ij,s} = \frac{E_j^+}{E_i^+} = \frac{2n_i cos(\vartheta_i)}{n_i cos(\vartheta_i) + n_j cos(\vartheta_j)} \ . \tag{SI 10}$$

Again, this can be summarized in the matrix form with the interface matrix $T_{ij}$. The later form of $T_{ij}$ follows directly from the Fresnel equations by $r_{ij} = -r_{ji}$.

$$\begin{pmatrix} E_i^+(z) \\ E_i^-(z) \end{pmatrix} = \frac{1}{t_{ij}} \begin{pmatrix} 1 & r_{ij} \\ r_{ij} & 1 \end{pmatrix} \begin{pmatrix} E_j^+(z) \\ E_j^-(z) \end{pmatrix} = T_{ij} \begin{pmatrix} E_j^+(z) \\ E_j^-(z) \end{pmatrix}. \tag{SI 11}$$

Now, the complete structure can be modelled layer-by-layer by multiplication of the corresponding phase and interface matrices,

$$T_{structure} = T_{01}P_1 \dots P_{n-1}T_{(n-1)n}. \tag{SI 12}$$

If the light incidents only from the left side of the structure ($E_n^- = 0$), the reflection and transmission coefficient of the complete structure follows the matrix elements $s_{ij}$ of $T_{structure}$,

$$r_{0n} = \frac{E_0^-}{E_0^+} = \frac{s_{21}}{s_{11}} \ , \tag{SI 13}$$

$$t_{0n} = \frac{E_n^+}{E_0^+} = \frac{1}{s_{11}} \ . \tag{SI 14}$$

With this the reflection and transmission of the incident light is obtained.

$$R_{0n} = |r_{0n}|^2, \tag{SI 15}$$

$$T_{0n,s} = \frac{\Re(n_n cos(\vartheta_n))}{\Re(n_0 cos(\vartheta_0))}|t_{0n}|^2. \tag{SI 16}$$

Waves of different wavelength are simulated to obtain a spectrum. The waves' angle of incidence is varied for angle-resolved spectra.

To derive the field at position $z$, the right as well as left propagating wave at $z$ are determined. The field at a layer interface serves as a starting point. Positions inside of one layer follow directly by using the phase matrix. Transfer coefficients $\tau_{ij}$ are defined as connection between the incident field $E_0^+$ with the inner field amplitude at a layer's interface ($E_i^+$ and $E_i^-$).

$$\frac{E_i(z)}{E_0^+} = E_i^+(z) + E_i^-(z) = \frac{E_i^+}{E_0^+}exp(-\varphi_i(z)) + \frac{E_i^-}{E_0^+}exp(\varphi_i(z)) = \tau_{0i,+}exp(-\varphi_i(z)) + \tau_{0i,-}exp(\varphi_i(z)) \ . \tag{SI 17}$$

These coefficients can be expressed in terms of the transfer matrix $T^I = T_{0i}$ and $T^{II} = T_{in}$ with their matrix elements $s_{ij}^I$ and $s_{ij}^{II}$, respectively.

$$\tau_{0i,+} = \frac{E_i^+}{E_0^+} = \frac{1}{det(T^I)}(s_{22}^I - r_{0n}s_{12}^I), \tag{SI 18}$$

$$\tau_{0i,-} = \frac{E_i^-}{E_0^+} = \frac{1}{det(T^I)}(r_{0n}s_{11}^I - s_{21}^I). \tag{SI 19}$$

The reflection coefficient $r_{0n}$ can be expressed as well by $s_{ij}^I$ and $s_{ij}^{II}$.

$$r_{0n} = \frac{E_0^-}{E_0^+} = \frac{s_{21}}{s_{11}} = \frac{s_{21}^I s_{11}^{II} exp(i\varphi_i(z)) + s_{22}^I s_{21}^{II} exp(-i\varphi_i(z))}{s_{11}^I s_{11}^{II} exp(i\varphi_i(z)) + s_{12}^I s_{21}^{II} exp(-i\varphi_i(z))}. \tag{SI 20}$$

Based on this, the following equation can be found to model the field at position $z$,

$$E_i(z) = \frac{s_{11}^{II} exp(-i\varphi(d_i-z)) + s_{21}^{II} exp(i\varphi(d_i-z))}{s_{11}^I s_{11}^{II} exp(i\varphi(d_i)) + s_{12}^I s_{21}^{II} exp(-i\varphi(d_i))} E_0^+. \tag{SI 21}$$

## Phase-shift in a planar open microcavity with a spacing layer

Here, we further explain the phase shift inside of the cavity in order to give an expression to estimate the necessary angular change for a given cavity length change. The phase shift of the complete cavity consists of the phase shift at each mirror surface $\varphi_{mirror}$, the phase shift inside of the solid spacer layer (here PMMA) and the air layer, $\varphi_{PMMA}$ and $\varphi_{Air}$, respectively.

$$\varphi_{cav} = \varphi_{mirror,1} + \varphi_{mirror,2} + \varphi_{PMMA} + \varphi_{Air}. \tag{SI 22}$$

In order to keep the cavity resonant to the exciton, this phase shift must remain constant for each mirror separation. Therefore, the change of the phase shift $\Delta\varphi_{cav}$ should be a multiple of $2\pi$. For simplicity, the phase shift change at the mirror surface for a tuned angle of incidence is neglected in the following.

$$m\,2\pi = \Delta\varphi_{cav} = (\varphi_{PMMA}(L_{cav,1}) + \varphi_{Air}(L_{cav,1})) - (\varphi_{PMMA}(L_{cav,2}) + \varphi_{Air}(L_{cav,2})) =$$

$$= (n_{PMMA}d_{PMMA}cos(\vartheta_1) + n_{Air}d_{Air}cos(\vartheta_1)) - (n_{PMMA}d_{PMMA}cos(\vartheta_2) + n_{Air}(d_{Air} + \Delta z)cos(\vartheta_2)),$$

$$with\ m\epsilon\mathbb{N}\ . \tag{SI 23}$$

The connection between the current angle of incidence $\vartheta_2$ and the change of the cavity length $\Delta z$ is thus found by including the expression for the phase shift, $\varphi_i = 2\pi n_i d_i \cos(\vartheta_i)/\lambda$,

$$cos(\vartheta_2) = \frac{n_{PMMA}d_{PMMA} + n_{Air}d_{Air,1}}{n_{PMMA}d_{PMMA} + n_{Air}(d_{Air,1} + \Delta z)}cos(\vartheta_1). \tag{SI 24}$$

Thus, the connection between the change of the cavity length and the resonant angle can be mainly addressed through the choice of the thickness of the spacing layer and the air gap. The resonant angle differs for higher modes from that of the cavity's ground state (Fig. SI 3).

## Connection between air gap and the resonant angle for different modes

The connection between the resonant angle of incidence $\theta_{res}$ ($E_{cav}$ ($\theta_{res}$) - $E_{ex}$ = 0) and the open-cavity spacing $\Delta z$ according to Equation (SI 24) is shown in Fig. SI.3 (a) for a 178.2 nm thick PMMA spacing layer. For comparison, the equivalent simulated relationship based on the transfer-matrix method is shown in Fig. SI.3b. For clarity it is noted that when PMMA is involved, the microcavity structure type corresponds to that of Fig. 1b. In Fig. SI.3 both graphs display the trend for the first resonant mode (blue solid line/triangles) of the cavity and the next higher mode (green dashed line/circles). The open-cavity spacing for the two consecutive resonator modes at normal incidence is thereby 48 nm and 356.5 nm, respectively. The phase shift change due to the angle-dependent reflectivity of the mirrors (and thereby the angle dependency of $\varphi_{mirror}$) is neglected in Equation (SI 24) and leads to the differences between the two plots.

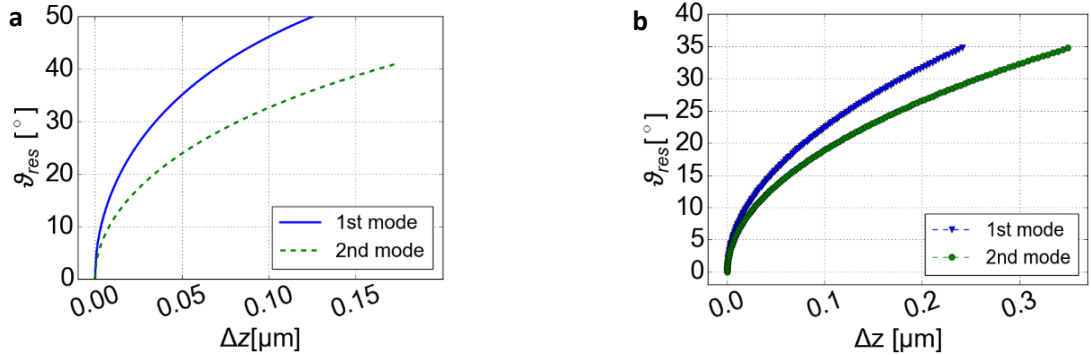

**Fig. SI. 3**: (a) Resonant angle of incidence as a function of the spacing distance $\Delta z$ for the fundamental and second resonator mode according to Eq. (SI 22). (b) Equivalent trends based on the transfer-matrix-method simulation for a representative configuration.

### Relative field strength for higher resonant modes

To show the influence of the mode number on the standing-wave pattern at resonance conditions for three consecutive modes, i.e. the fundamental resonator modes and the two higher ones, the calculated relative field strength as well as the angle-resolved reflectivity spectra are plotted in Fig. SI.4. The simulations were performed for the case of an 178.2 nm thick PMMA layer on top of the left DBR and a resonant angle of incidence $\vartheta_{res}$ of 5° corresponding to air gaps for the three different modes of 51.0 nm, 360.8 nm and 670.5 nm, respectively. Here, similar to the considerations regarding the graphs in Fig. 4, only the reflectivity of the 2D-material is considered but not its absorption (the result in (b) corresponds to that in Fig. 4a shown for the second resonator mode). The splitting within the reflectivity spectra is smaller for higher modes due to the smaller mode volume.

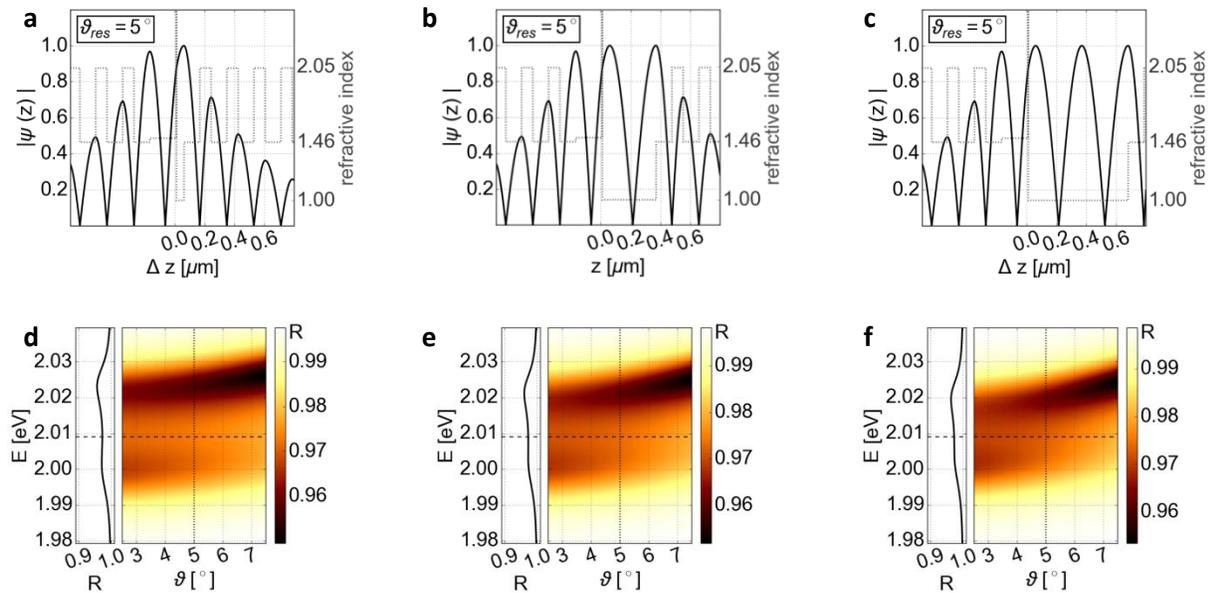

**Fig. SI. 4**: Theoretical relative field strength for the first cavity mode (a), the second (b) and the third resonator mode (c) for an open microcavity structure with an angle of incidence of 5° corresponding to that in Fig. 1b. The PMMA layer on top of the left DBR is 178.2 nm thick and the air gaps are 51.0 nm,

360.8 nm and 670.5 nm for the first three modes, respectively. (d-f) show the corresponding calculated angle-dependent reflectivity for the different cases similar to Fig. 4.

**Comparison of the coupling situation for two resonant modes**

For the ground state and the energetically next higher cavity mode, the relationship between the cavity's length change and the necessary resonant angle of incidence $\vartheta_{res}$ differs because the air gap of the higher mode is larger (Equation SI 1, Fig. SI. 2). Thus, for a higher mode, a larger change of the mirror separation is necessary to reach the same shift for the resonant angle. On the other hand, the same angular change leads to the same modification of the relative field strength at the exciton position. Altogether, the splitting in our simulation is higher for the first resonant mode due to the smaller mode volume (Fig. SI.5).

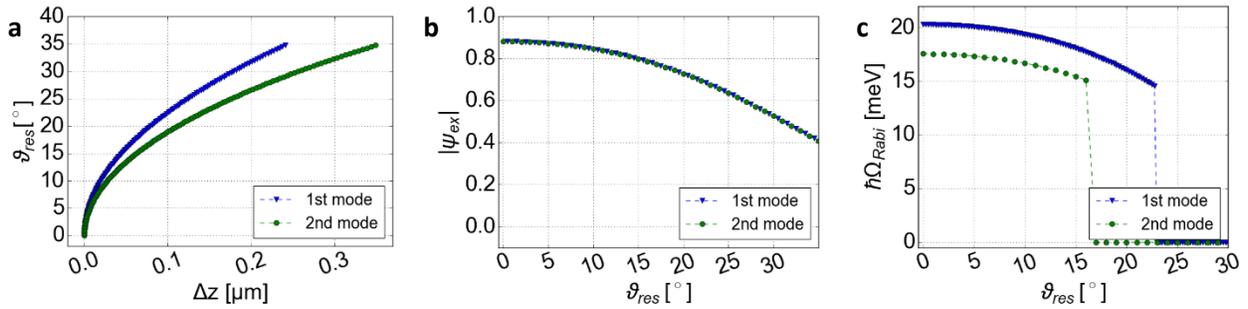

**Fig. SI.5**: Relationship between the cavity length change $\Delta z$ and the resonant angle $\vartheta_{res}$ (a), the relative field strength (b) and the resolvable splitting $\hbar\Omega_{Rabi}$ within the reflectivity spectrum (c) for the fundamental resonant cavity mode as well as the energetically next higher mode for a structure with a 178.2 nm thick PMMA spacing layer. In (b) and (c), $\vartheta_{res}$ is directly linked with the corresponding $\Delta z$ in (a).

**Determination of the resonant angle**

The resonant angle at a given cavity length is determined iteratively until spectral overlap between the cavity mode and the exciton's emission wavelength within the calculated reflectivity spectrum is reached, while the absorption of $WS_2$ is neglected (representing an uncoupled situation). For a 178.2 nm thick PMMA spacer layer and an air gap of 356.5 nm, this is the case for normal incidence shown in Fig. SI.6a. The split between the normal modes in the calculated reflectivity with included absorption of $WS_2$, which represents the spectrum in the coupled situation, is slightly asymmetric, as can be seen in Fig. SI.6b, although the "empty" (uncoupled) cavity mode is energetically resonant with the exciton energy.

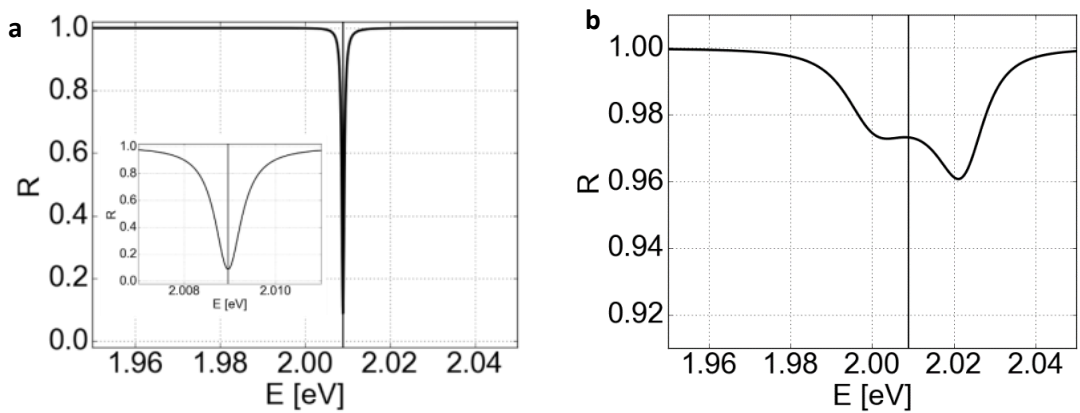

**Fig. SI. 6**: Calculated reflectivity spectrum for an uncoupled cavity with inserted WS$_2$ (a). Inset: close-up spectrum of the resonance. In comparison, the coupled cavity shows a mode splitting in the calculated reflection spectrum with included absorption of WS$_2$ (b). The slight asymmetry at determined resonance condition is attributed to the role of the absorbing layer on the actual cavity field configuration, which is approximated for the microcavity system when neglecting the imaginary part of the dielectric function. The extracted resonance energy from (a) is indicated by the vertical line in both graphs.

## Angle resolved reflectivity of the DBR mirror

The angle-resolved reflectivity of a SiO$_2$/Si$_3$N$_4$ DBR with 11.5 layer pairs according to transfer-matrix-method calculations is depicted in Fig. SI.7a. Over the explored range of angles (smaller than 40° in this study) the reflectivity change as a function of incidence angle is less than 0.1%, which will have a minor impact on the cavity Q-factor for different angles. Fig. SI.7 shows that the field distribution at normal incidence (b) has a maximum at the DBR's surface, whereas for an angle of 35° this maximum is shifted (c). Thus, the DBR contributes an additional phase shift at higher angles of incidence.

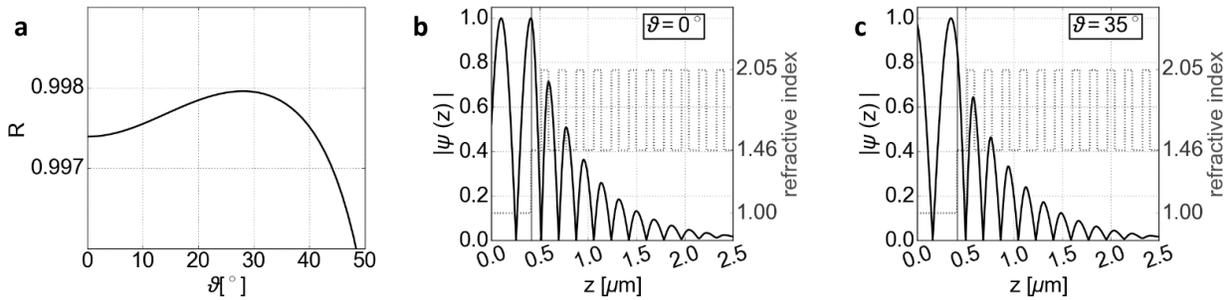

**Fig. SI.7**: (a) Calculated DBR reflectivity as a function of the incidence angle. For comparison, the field distribution in the vicinity of the DBR surface is shown for normal incidence (b) and a 35 degree incidence angle (c).

## Angle resolved transmission and absorption spectra for s-polarized light

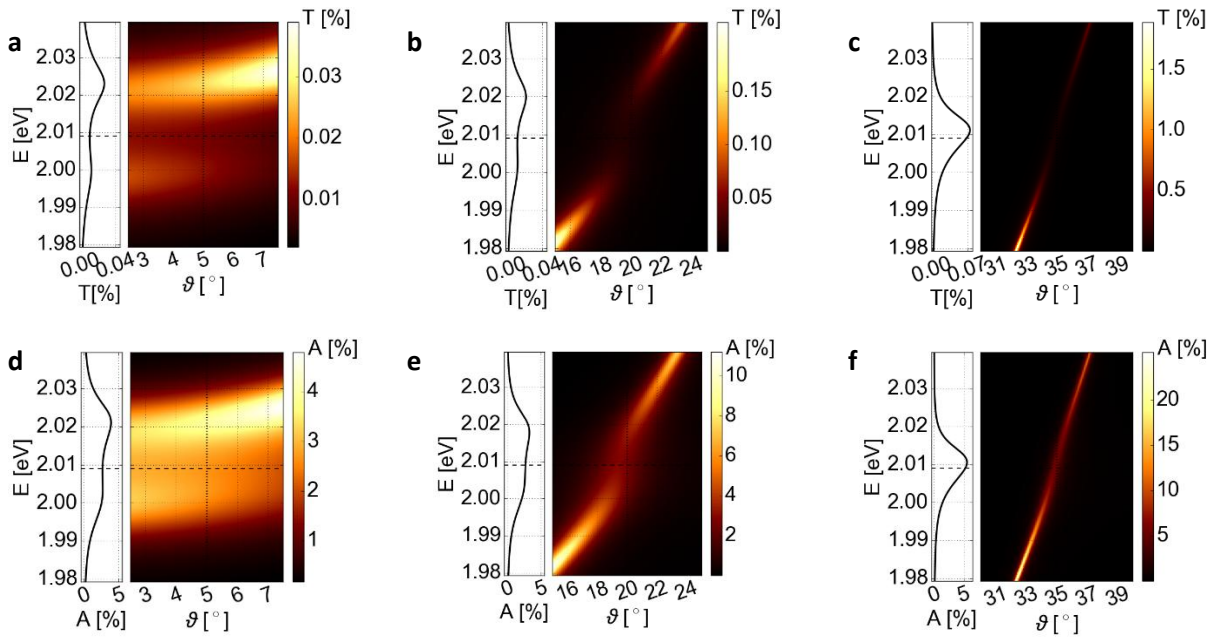

**Fig. SI.8**: Influence on the transmission (upper row) and absorption (lower row) for s-polarized light by a tuned cavity length. The dashed vertical line marks the exciton's emission energy, which is identified by the peak position within the imaginary part of the refractive index. Within the absorption spectra, the splitting is nearly symmetric. As expected from literature, the splitting in reflection, transmission and absorption differs [S6]. Note, that the observed splittings depend on the cavity's as well as exciton's linewidth.

For the reflectivity calculations provided in Fig. 4, corresponding angle-resolved transmission and absorption spectra are summarized in Fig. SI.8. For the microcavity structure type of Fig. 1b, the spacer layer amounts to 178.2 nm in the calculated examples. For the structure's air gap of 360.8 nm, 425.6 nm and 575.7 nm, the resonant angle lies at 5°, 20° and 35°, respectively.

**Angle resolved reflectivity, transmission and absorption spectra for p-polarized light**

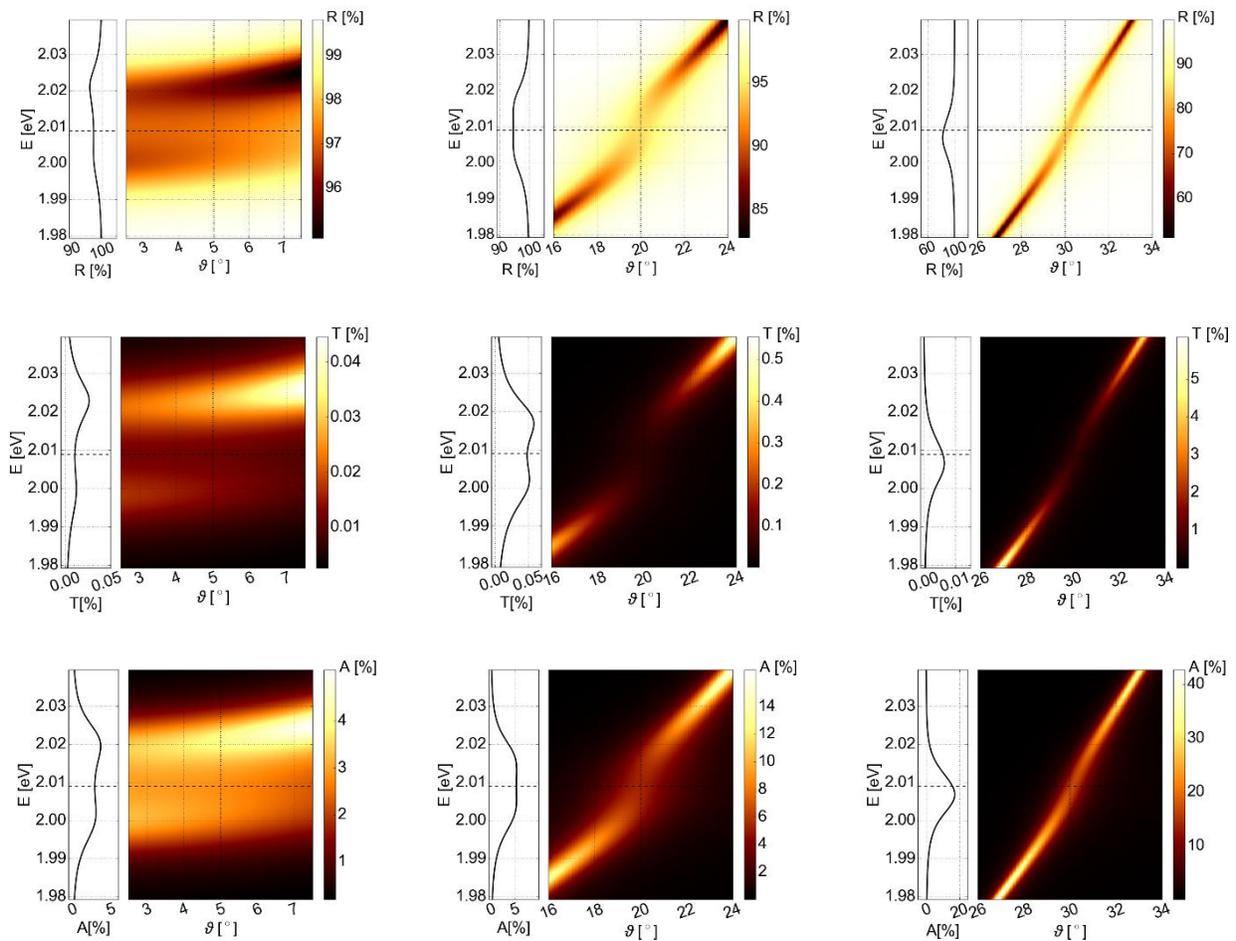

**Fig. SI.9**: Impact on the reflection (upper row), transmission (middle row) and absorption (lower row) for p-polarized light by a tuned cavity length. The dashed vertical line marks the exciton's emission energy, which is identified by the peak position within the imaginary part of the refractive index.

Figure SI.9 shows the influence on the transmission (upper row) and absorption (lower row) for p-polarised light of a tuned cavity length with a 178.2 nm thick PMMA spacer layer on one mirror (structure depicted in Fig.1b). The structure's air gap is 360.8 nm, 425.6 nm and 515.0 nm with a resonant angle at 5.08°, 20.315° and 35.008°, respectively. The air gap corresponds thereby to resonant angles for s-

polarized light of 5°, 20° and 35°, respectively. Thus, the determined resonant angles for s and p polarized light have only a small deviation from each other. The exciton's emission energy is represented by the dashed vertical line. The spectral detuning of the cavity mode by the exciton's absorption differs compared to s-polarized light. The overall effect of a continuously tunable coupling is nevertheless also observable for p-polarized light. Unpolarized light will lead to a weighted sum of the p- and s-polarized spectra.

## Supporting Information References